# Exploring Deep Learning Techniques for Glaucoma Detection: A Comprehensive Review

Aized Amin Soofi
*Department of Computer Science*
*National University of Modern Languages*
Faisalabad, Pakistan
aizedamin@gmail.com

Fazal-e-Amin
*Department of Software Engineering*
*King Saud University*
*College of Computer and Information Sciences*
Riyadh, KSA
fazal.e.amin@gmail.com

*Abstract*—Glaucoma is one of the primary causes of vision loss around the world, necessitating accurate and efficient detection methods. Traditional manual detection approaches have limitations in terms of cost, time, and subjectivity. Recent developments in deep learning approaches demonstrate potential in automating glaucoma detection by detecting relevant features from retinal fundus images. This article provides a comprehensive overview of cutting-edge deep learning methods used for the segmentation, classification, and detection of glaucoma. By analyzing recent studies, the effectiveness and limitations of these techniques are evaluated, key findings are highlighted, and potential areas for further research are identified. The use of deep learning algorithms may significantly improve the efficacy, usefulness, and accuracy of glaucoma detection. The findings from this research contribute to the ongoing advancements in automated glaucoma detection and have implications for improving patient outcomes and reducing the global burden of glaucoma.

*Index Terms*—Glaucoma Detection, Deep Learning Techniques, Medical Imaging, Glaucoma Segmentation, Comprehensive Survey, Machine Learning

## I. INTRODUCTION

The second most common condition in the world that results in permanent nerve damage to the eyes is glaucoma, which can impair vision completely or partially. Glaucoma is known as the" silent thief of sight" because it often remains untreated in its early stages [1]. Studies have shown that glaucoma remains a significant global health concern. Over 60 million instances of glaucoma are thought to exist globally, with Open-Angle Glaucoma (OAG) involving on average 1.69% of people and Angle-Closure Glaucoma (ACG) 0.69% [2]. Additionally, this chronic eye condition affects about 3.54% of people worldwide between the ages of 40 and 80. Glaucoma affects over 4.5 million people worldwide and is a contributing factor in 8% of blindness [3]. Notably, in Southeast Asia alone, approximately 13.6% of the population is affected by glaucoma.

By the year 2025, the number of individuals worldwide suffering from glaucoma is projected to reach 80 million, signifying a substantial increase over the past decade. Within 10 years, the combined figures of OAG and ACG are expected to rise by 20 million people. Specifically, Asia is anticipated to experience an 87.6% increase in ACG cases, while approximately 58.6 million individuals will be affected by OAG. Also, the total no. of patients will be 111.8M in 2040 [2]. These statistics paint a concerning picture of the global prevalence of glaucoma, highlighting the urgency to enhance detection methods. Deep learning techniques present a promising avenue for automated glaucoma detection, which can contribute to addressing the growing burden of this eye disease [4]. Table I shows the statistics of global affection of glaucoma.

An understanding of glaucoma is essential for developing effective detection methods [5]. The human eye's anatomy and fluid dynamics play a crucial role in maintaining eye health. Aqueous humor, a transparent fluid located at the front of the lens and behind the cornea, is vital for maintaining Intraocular Pressure (IOP) and nourishing eye structures [6]. Under normal conditions, the formation and drainage of aqueous humor are balanced, ensuring a healthy IOP. However, in glaucoma, the drainage system becomes compromised, leading to the accumulation of fluid in the anterior chamber. This fluid buildup increases the IOP, which can cause damage to optic nerves, visual fields, and the development of optic cupping, a common sign of glaucoma [7]. The representation of healthy and glaucoma-affected OD and OC is shown in figure 2.

In glaucoma, higher IOP can cause optic neuropathy and worsen abnormal cupping of the Optic Disc (OD). The Cup-to-Disk ratio (CDR) compares the diameter of the cup to the total diameter of the disc to assess the severity of glaucoma [8]. The CDR indicates the progression of glaucoma, with a normal CDR being around 0.5 [9]. Glaucoma affects a significant number of people worldwide, with projections indicating a substantial increase in cases. Two prevalent forms of glaucoma—OAG and ACG—have different symptoms and progression of the disease. Early detection of glaucoma is challenging as most types exhibit no early symptoms, making it crucial to develop accurate and efficient detection methods [10].

Traditional manual detection methods, such as Tonometry, assessment of Optic Nerve Head (ONH), and visual field tests, have limitations in terms of cost, time, and accuracy [11]. Recent advancements in computer science have paved the way for Computer-Aided Detection (CAD) tools, utilizing machine learning techniques such as Support Vector Machines (SVM), Fourier Transformation (FT), Wavelet Transformation (WT), and K-Nearest Neighbor (KNN). However, these approaches rely on manual feature extraction, which involves human in-tervention to identify relevant characteristics in retinal fundus images [12]. This process is subjective, as different experts may extract distinctive features, leading to inconsistency and potential bias in the results. Additionally, it takes a lot of time to manually extract features, requiring considerable attempt and expertise. Furthermore, the handcrafted features may not fully capture the intricate patterns present in retinal fundus images. The complexity of glaucoma-related features and variations in the images pose challenges for manual feature engineering [13]. Human experts may overlook or fail to represent all the relevant information necessary for accurate glaucoma detection, limiting the effectiveness of these machine learning methods [14].

TABLE I
GLOBAL GLAUCOMA: A GROWING CONCERN - STATISTICS AND PROJECTIONS

| Year | Statistics | Percentage/Number |
|---|---|---|
| 2010 | Total global glaucoma cases (estimated) | >60 million |
| 2010 | Open-Angle Glaucoma (OAG) prevalence | 1.69% of the population |
| 2010 | Angle-Closure Glaucoma (ACG) prevalence | 0.69% of the population |
| 2010 | Glaucoma prevalence in Southeast Asia | 13.6% of the population |
| 2020 | Estimated global glaucoma cases | 80 million |
| 2010-20 | Increase in OAG and ACG cases over 10 years | 20 million |
| 2020-40 | Estimated increase in ACG cases in Asia | 87.60% |
| 2020-40 | Estimated number of individuals affected by OAG | 58.6 million |
| 2040 | Projected global glaucoma cases | 111.8 million |

Deep learning techniques have recently become more popular in the medical and healthcare fields partly due to their ability to learn autonomously from data and extract relevant features. Deep neural networks are used in deep learning, a branch of machine learning, to carry out a variety of operations, such as image segmentation, classification, and detection. Among the designs used in deep learning are Convolutional Neural Networks (CNN), Recurrent Neural Networks (RNN), and Long Short-Term Memory (LSTM) [15].

*A. Motivation*

This study intends to investigate the use of deep learning methods for glaucoma detection and available datasets for glaucoma detection. These methods eliminate the need for human feature extraction by automatically extracting pertinent information from photos or datasets by using deep neural networks. By increasing precision, effectiveness, and early diagnosis, the use of deep learning in glaucoma detection has the potential to revolutionize the industry. The hierarchical representation of deep learning techniques is shown in figure3. The research taxonomy is presented in figure 1.

*B. Contribution*

The goal of this research investigation is to present an in-depth overview of the cutting-edge deep-learning methods utilized in the segmentation, classification, and detection of glaucoma. By reviewing recent studies from 2019 to 2023, we aim to analyze the effectiveness and limitations of these techniques, highlight key findings, and identify potential areas for further research. It also aims to contribute to the ongoing advancements in automated glaucoma detection, improving the accuracy, accessibility, and efficiency of glaucoma diagnosis.

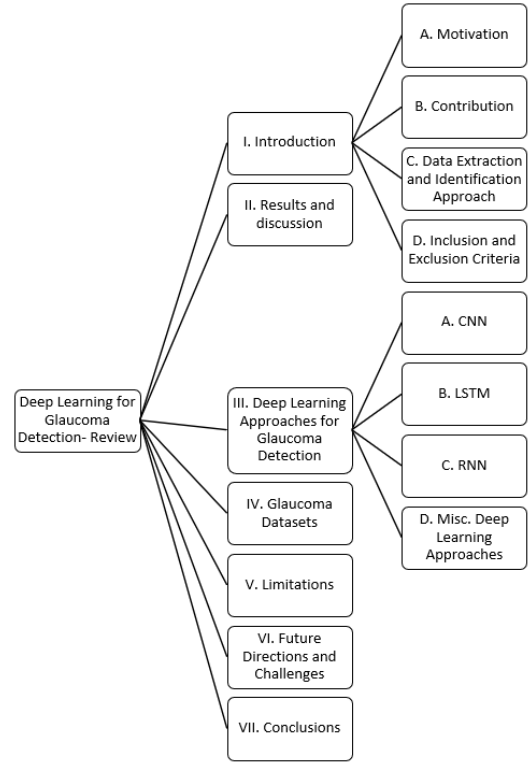

Fig. 1. Research Taxonomy

*C. Data Extraction and Identification Approach*

The following electronic databases were taken into account in this review: PubMed, IEEE Xplore Digital Library, Science Direct, Springer, Scopus, and Google Scholar. All the articles were printed between the beginning of 2019 and the middle of 2023 in various publications. The keywords used to extract the required studies to narrow the research include; "Deep Learning And Retinal Image", "Deep Learning And Glaucoma", "Glaucoma AND machine learning", "Deep Learning And Machine Learning For Segmenting Glaucoma", and "Deep Learning And Glaucoma segmentation". A total of 13,324 results were retrieved from the keyword search. When the criteria for rejection were applied, 114 works were left. The results of initial scanning from selected databases have been presented in figure 4.

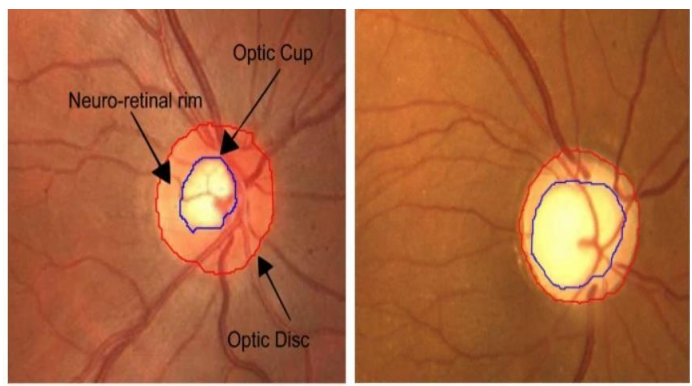

Fig. 2. Representation of healthy and glaucoma-affected Optic Disk and Optic Cup

*D. Inclusion and Exclusion Criteria*

The following factors were disregarded throughout the selection process:
- Papers that did not have the words "Glaucoma" and "Deep Learning" in the title.
- Articles that did not use the terms "Glaucoma" and "Deep Learning" in their keywords.
- Articles that did not have the term "Glaucoma" in their metadata.

Selective measures included a review of each study's abstract, keywords, and methodology. This allowed for the acquisition of the studies' selection criteria. The remaining 114 studies' abstracts, keywords, and methodologies were evaluated for analysis of their significance and impact. At this point, the most relevant studies were chosen by applying the criteria that include:
- Data collection via retinal imaging.
- Processing methods involving deep learning.
- Analysis of OD structure.
- Imaging classification procedures.
- Impact factor and citation info for the journal and article.

Using these standards, we chose 52 papers for a comprehensive review. Thus, studies that only presented segmentation and not classification were disregarded. In summary, 31 publications were considered that dealt with deep learning techniques for glaucoma detection. The year-wise results of selected studies are shown in figure5. This research is organized as follows: The results and discussions of the conducted study have been presented in section II. A review of the literature regarding deep learning approaches for glaucoma detection is included in section III. The datasets used to identify glaucoma are presented in section IV. The study's limitations are detailed in section V. Future directions and challenges have been presented in section VI and section VII summarizes the major contributions and emphasizes the importance of deep learning approaches in glaucoma diagnosis to bring the article to a close.

## II. RESULTS AND DISCUSSION

The provided studies address various problems related to glaucoma detection and diagnosis. Some of the addressed problems include improving the accuracy and efficiency of glaucoma detection, segmentation of key structures such as the optic disc and optic cup, classification of glaucoma stages, detection of initial and advanced glaucoma, reducing redundancy in fundus images, evaluating the performance of different models, predicting glaucoma progression, predicting future visual field data, distinguishing glaucoma from other eye conditions, reducing computational intensity for real-time applications, early detection of glaucoma, automated diagnosis of ocular diseases, segmenting optic disc images, identifying multiple diseases causing glaucoma, improving segmentation accuracy, and reducing redundancy and effectiveness issues in glaucoma detection models.

These studies employ various techniques such as CNN models, RNN models, LSTM models, hybrid models, segmentation algorithms, feature extraction, classification models, and optimization techniques to address these problems. The results of the review show that CNN is the most prominent approach in the detection of glaucoma. CNN can pull out spatial details from pictures. Most of the time, retinal images or scans of the optic nerve are used to find glaucoma. CNNs are great at finding important local patterns, textures, and structures that point to changes caused by glaucoma. Because of this, CNNs are a good way to study the complex visual features of the retina.

The combination of both LSTM and CNN has also gained attraction from the research community in the detection of glaucoma disease. When LSTM and CNN are used together, they make a powerful framework for detecting glaucoma. This is because the model can catch both the spatial and temporal aspects of the disease. By taking advantage of these benefits, a combination of LSTM and CNN models can help researchers and doctors better understand how glaucoma gets worse and how it can be treated. RNN is the least used approach for glaucoma detection because it can have the "vanishing gradient" or "exploding gradient" problem, which makes it hard to get and use useful information to find glaucoma. Figure 6 shows the different deep-learning techniques that are in use for glaucoma detection. Table II, III, IV, and V are the abstract level representation of all the selected previous work in terms of methodology, problem, strengths, and weaknesses.

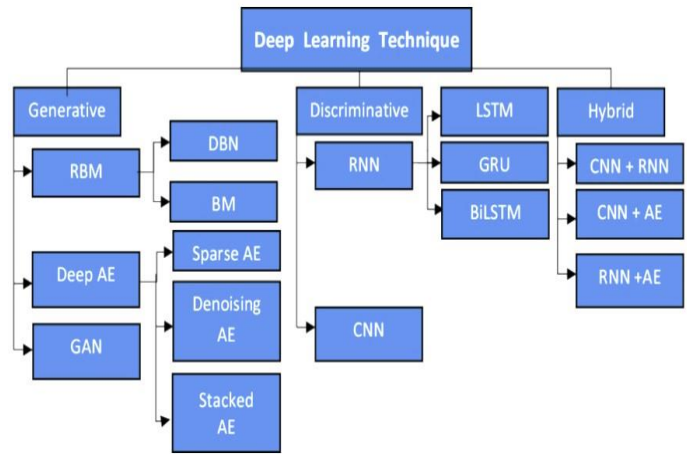

Fig. 3. The Hierarchical representation of deep learning techniques

In the field of medical image analysis for glaucoma detection, several studies have utilized different methodologies to achieve accurate results. In [16] employed U-Net Mobile-Net V2 and achieved an accuracy of 0.88, sensitivity of 0.91, specificity of 0.86, and an area under curve (AUC) of 0.93. The study in [17] explored multiple methodologies and found that DENet had the highest accuracy and AUC, VGG19 had the highest sensitivity, ResNet had the highest specificity, and GoogleNet had the highest F1 score. An AG-CNN [18] was used to achieved high accuracy, sensitivity, specificity, and AUC values. To achieve moderate accuracy values, ResNet and GoogleNet was applied in [19]. Other studies employed different techniques such as Inter-GD, Joint-RCNN, CNN and RNN, DenseNet, SD-OCT-based CNN, CNN-LSTM, AED-HSR, Joint Segmentation, Deep Neurofuzzy Network, ResNet-50 CNN, GlauNet CNN, RNN, CNN (Inception V3, DenseNet, ResNet), modified M-Net with BiConvLSTM, and BiDCU-Net, each achieving varying levels of accuracy, sensitivity, specificity, and other metrics. Overall, the performance of these methodologies highlights the potential for accurate glaucoma detection. In table VI, we represent the results of each model in context to its accuracy (Acc), sensitivity (SN), specificity (SP), AUC, and measure of f-score. It is evident that the hybrid models like CNN-RNN, AG-CNN, CNN-LSTM, AED-HSR approach, SD-OCT-based CNN, GlauNet CNN, Modified M-Net with BiConvLSTM and BiDCU-Net [19]–[26] provide state of art results. Table VI also presents the results and comparison of literature that we considered in this study.

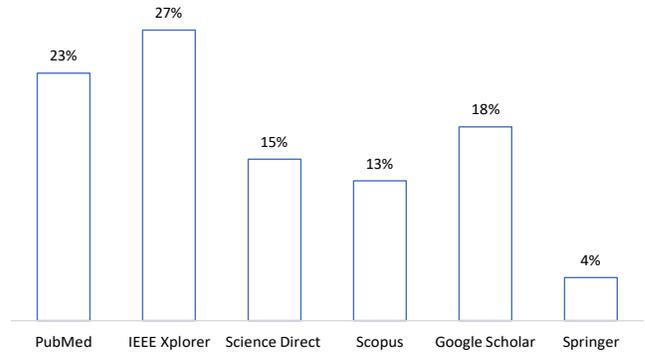

Fig. 4. Results of initial scanning from selected databases

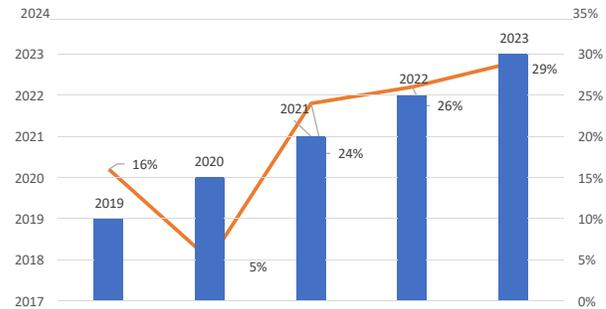

Fig. 5. Year-wise results of included studies

## III. DEEP LEARNING APPROACHES FOR GLAUCOMA DETECTION

Deep learning methods are crucial in detecting glaucoma because they can offer timely and precise diagnoses, promote healthcare accessibility, and improve the efficiency of healthcare systems. These advancements have the capacity to significantly influence the well-being of individuals who are susceptible to glaucoma and the overall public health scenario. In this section different deep learning approaches for glaucoma detection have been explored.

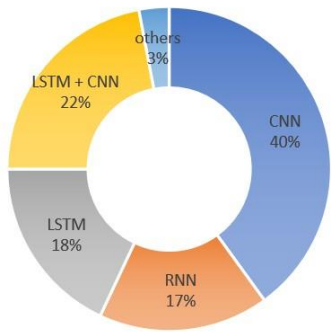

Fig. 6. Use of different deep learning techniques for glaucoma detection

*A. CNN*

The problem addressed in [16] lies in the need to improve the accuracy and efficiency of glaucoma detection and proposed a model that enhances accuracy and efficiency. Traditional methods often face limitations in accurately identifying key structures, such as the OD and optic cup (OC), in retinal images. To overcome this, the proposed solution combines segmentation and classification using CNN. In the segmentation stage, U-Net is utilized for precise feature extraction of the OD and OC. This involves various processes like pre-processing, data augmentation, and training on datasets like RIM-ONE and DRISHTI. Subsequently, a classical CNN architecture based on the MobileNet V2 model is employed for the direct classification of glaucoma in the second stage. By integrating the results from both stages, a comprehensive report is generated, aiding ophthalmologists in diagnosis.

The study [28] focused on the challenges of detecting glaucoma. Manual searching of ROI creates noise thus accuracy gets affected. Also, too much global information is lost in process of cropping the images for finding ROI. To overcome this, the author put up a model with two basic components. First, OD and OC were divided up using U-Net CNN. It also uses an autoencoder decoder with a 5-layer CNN, of which 2 layers contain the ReLu function. The OD and OC are estimated individually by two heads. Then, to evaluate CRD, the OD and OC data are combined. The second step is the CDR prediction based on the evaluation. In previous studies, it has always been thought of as two different procedures to segment the OD and OC when detecting glaucoma.

Researchers have focused on developing individual segmentation models for OD and OC, aiming to accurately delineate these structures in retinal images [28], but in [30] work on a different approach for this problem had been performed. While their work was more focused on developing a joint segmentation model for OD and OC. The proposed method is divided into four stages. At first, VGG16 architecture with atrous layers is used for feature extraction. The only change made is replacing the last three layers with 9 convolutional layers. Instead of using pixel-based segmentation, object detection is used. The major reason for using the atrous layer is to improve the accuracy of boundary boxes. The second and third stages are of OD and OC proposal region detection. Two faster RCNN models are used for this. Feature extracted from previous stages given as input to it. For the OD proposal network, ROI is used to crop the mapped features. The highest confidence is found for the disk proposal network. Finally, the results are fused through Disk-attention based module.

The problem addressed in [18] is the presence of large redundancy regions in fundus images, which can negatively impact the performance of glaucoma detection models. These redundant regions can mislead CNN models by diverting their focus to irrelevant information. To overcome this problem, the authors provide a framework that incorporates an attention mechanism with the CNN model. Three primary elements make up the Attention-based Glaucoma CNN: attention prediction, localization of problematic regions, and glaucoma categorization. Utilizing the 11,760 labelled fundus images in the LAG database, the model undergoes training and assessed. 5,824 of these photos, which are acquired from four ophthalmologists, are chosen for use in creating attention maps for the identification of glaucoma.

CNN models used in glaucoma detection often face challenges in terms of performance. This problem was acknowledged in [17], which sought to assess the SP, SN, and Acc of several pre-trained CNN models, including Standard CNN, VGG19, ResNet, DENet, and GoogleNet. They conducted their study using the RIM-ONE, DRISHTI, and ESPERANZA datasets. The study revealed that VGG19, when combined with transfer learning and fine-tuning techniques, exhibited superior performance compared to other models. However, the researchers didn't stop there. They further investigated the integration of patient clinical history data to enhance the glaucoma detection models' accuracy and effectiveness. The investigation took into account a number of variables, including age, IOP, and family history. Gomez et al. wanted to enhance the overall accuracy of CNN-based glaucoma detection, making a step towards more precise and dependable diagnostic tools by including these clinical aspects into the models. To address the time-consuming issue of previous techniques, [34] also use CNN model with preprocessing techniques to build a non-invasive way for detecting glaucoma at early stages. CNNs offer advantages such as minimal preprocessing and the ability to update weights in the hidden layers. By following the traditional CNN architecture and incorporating convolutional and pooling layers along with activation functions like ReLU and sigmoid, the proposed model preprocesses and classifies the retinal image, enabling the early detection of glaucoma.

A similar task using CNN approach was performed in [36] in which CNN Residual Network-34 was used to categorize glaucoma using photos of the fundus. The study investigates the effects of three optimization techniques (Adam, SGD, and RMS) and the training-to-testing data ratio using a dataset of 2156 fundus pictures classified into early, moderate, deep, OHT, and normal glaucoma types. The results demonstrate an impressive 94% accuracy rate, showcasing the system's ability to accurately classify glaucoma based on fundus images and potentially alleviate the burden on manual CDR calculations by ophthalmologists. A CNN model [37] was used for early detection of glaucoma with pre-processing using Gaussian filtering to remove image noise. Following that, the Modified Level Set Algorithm is used to segment the OC. Derived morphological and non-morphological metrics include modified LBP, disc area, cup area, blood vessel properties, color, and shape. The Self Adaptive Butterfly Optimization Algorithm (SA-BOA) is used to optimize the weights of the CNN framework that is used to classify these characteristics. The suggested method achieves greater accuracy and smaller negative metrics, and the findings demonstrate notable gains when compared to traditional procedures.

TABLE II
SUMMARY OF CNN TECHNIQUES FOR GLAUCOMA DETECTION

| Ref. | Method | Problem | Strength | Weakness |
|---|---|---|---|---|
| [16] | Segmentation and Classification using CNN | Improving accuracy and efficiency of glaucoma detection | Integration of segmentation and classification | May not work well on other types of imaging for glaucoma detection [27] |
| [28] | U-Net, CNN, Autoencoder decoder, CDR prediction | Challenges in detecting glaucoma | Segmentation and CDR prediction with U-Net | Adopted models are hard to read, which makes it hard to understand how they make decisions [29] |
| [30] | VGG16, Atrous layers, Faster RCNN | Joint segmentation of OD and OC | Improved accuracy using object detection | Lead to less accurate results or more false positives or negatives if image quality varies [31] |
| [18] | Attention mechanism, CNN | Redundancy in fundus images affecting glaucoma detection | Combining attention mechanism with CNN | Tunning of additional hyperparameters for attention mechanism can degrade performance [32] |
| [17] | Evaluation of pre-trained CNN models, Integration of clinical data | Performance of pre-trained CNN models, Integration of clinical data | Comparative evaluation of CNN models | The link among clinical findings and glaucoma can be biased by differences in how diagnoses are made [33] |
| [34] | CNN with preprocessing | Time-consuming glaucoma detection | Early detection of glaucoma using minimal preprocessing | change some traits or information that are important for detecting disease [35] |
| [36] | CNN with ResNet-34 | Manual CDR calculations by ophthalmologists | Accurate classification of glaucoma based on fundus images | Fundus images depend on racial background, age, or the imaging tools used |
| [37] | CNN with preprocessing | Conventional techniques for glaucoma detection | Higher precision and improved metrics | High chances of changes in traits |
| [38] | ResNet-50 CNN | Reducing false positives and improving overall detection accuracy | Detection accuracy of 98.48% | High Training complexity |
| [39] | CNN-based Inception V3 | Efficient and effective glaucoma detection | Higher accuracy and AUC compared to other algorithms | input variations sensitivity |

A CNN model built on the ResNet-50 architecture was trained in [38] by using the grey channels of fundus images and data augmentation methods. The proposed model achieved impressive results, demonstrating a detection accuracy of 98.48%, a sensitivity of 99.30%, a specificity of 96.52%, an AUC of 97%, and an F1-score of 98% on the G1020 dataset. In [39] a dataset of 6072 fundus images were utilized, including 2336 glaucomatous and 3736 normal images. The 5460 images used for training and 612 for testing, the model achieved an impressive accuracy of 0.8529 and an AUC value of 0.9387. Notably, when compared to other algorithms such as DenseNet121 and ResNet50, the proposed model outperformed them with accuracies of 0.8153 and 0.7761, respectively. These findings highlight the potential of the CNN-based Inception V3 model as an efficient and effective approach for glaucoma detection, providing valuable insights for early diagnosis and improved patient management.

*B. LSTM*

Limitation of future visual field is another limitation that is faced in previous studies. In [40] suggested a deep learning algorithm to overcome this issue by using prior VF data and optical coherence tomography (OCT) pictures. The ResNet-50 model was utilized to analyze the image data, including thickness maps, vertical tomograms, and horizontal tomograms. A combination of image features and previous VFs was fed into a LSTM network for predicting future VF. The proposed model achieved a mean absolute error (MAE) of 3.31 ± 1.37 and a root mean square error (RMSE) of 4.58 ± 1.77" to maintain consistency. The integration of VF data and OCT images proved highly effective, and the model demonstrated utility in detecting and weighting noisy data.

Another study [41] addressed a similar issue. The approach involves fine-tuning the Efficientnet-b0 model for feature extraction, the Haar wavelet to reduce the number of features, and the Bi-LSTM to classify the results at the end.

TABLE III
SUMMARY OF LSTM TECHNIQUES FOR GLAUCOMA DETECTION

| Ref. | Method | Problem | Strength | Weakness |
|---|---|---|---|---|
| [40] | ResNet-50, LSTM | Predicting future visual field based on previous VF and OCT images | Integration of VF data and OCT images | Lack of interpretability |
| [41] | Fine-tuning Efficientnet-b0, Haar wavelet, Bi-LSTM | Accurate and efficient glaucoma stage classification | Higher classification accuracy with Bi-LSTM | Loss of fine-grained spatial information |
| [42] | LSTM-based CNN | Time-consuming diagnostic process | Automation and reduced diagnostic time | Trouble with setting [43] |
| [44] | Slide-level feature extractor, LSTM networks | Analyzing spatial dependencies in SD-OCT scans | Effective combination of features and LSTM | Glaucoma images show a loss of delicate spatial information |
| [25] | CNN-LSTM | Efficient and accurate ocular disease classification | High accuracy in distinguishing normal and cataractous cases | setting hyperparameters problem |
| [22] | Modified M-Net with BiConvLSTM | Accurate cup and disc segmentation for glaucoma detection | Remarkable performance in segmenting cup and disc regions | Require longer training time |

Evaluation of the model is performed using a large dataset of advanced glaucoma, early glaucoma, and normal images. Results demonstrate that the Bi-LSTM network achieves higher classification accuracy compared to other classifiers such as softmax, Support Vector Machine (SVM), and KNN. By leveraging deep learning techniques, this approach offers a promising solution for accurate and efficient glaucoma stage classification from fundus images. Another study [42] addressed a similar issue. LSTM-based CNN methods were introduced to automate the process and reduce diagnostic time. The proposed research explores CT image classification and novel applications such as COVID-19 detection and glaucoma detection using AI techniques. Predictions for future trends were also discussed. [44] introduced an innovative Time-aware Convolutional Long Short-Term Memory (TC-LSTM) unit to separate memories into short-term and long-term memories and utilize time intervals to penalise the short-term memory. The TC-LSTM unit is integrated into an auto-encoder-decoder architecture to enable the end-to-end model to effectively handle irregular sampling intervals of longitudinal GCIPL thickness map sequences. This allows the model to capture both spatial and temporal correlations. Empirical evidence demonstrates the superiority of the proposed model in comparison to the conventional approach.

A CNN-LSTM-based model architecture [25] was proposed for automated diagnosis of ocular diseases from fundus images, specifically distinguishing between normal and cataractous cases. The model trained on the ODIR dataset achieved an impressive accuracy of 97.53%, surpassing previous systems. This low-cost diagnostic system offers a promising solution for efficient and accurate ocular disease classification, contributing to timely interventions and improved patient outcomes. The challenges of accurate cup and disc segmentation was addressed in [22] for glaucoma detection through the development of a modified M-Net with bidirectional convolution LSTM. This powerful model combines encoder and decoder features, resulting in enhanced segmentation of cup and disc regions. When tested on REFUGE2 data, the model demonstrated remarkable performance, achieving a dice score of 0.92 for the optic disc and an impressive accuracy of 98.99% in segmenting cup and disc regions.

## C. RNN

Prediction of time progression is another factor that neglected in previous studies, it mostly primally focused on spatial data. Analyzing spatial correlations and collecting pertinent data from 3D spectral-domain optical coherence tomography (SD-OCT) images to identify glaucoma is difficult. In order to solve this, the researcher [45] suggested a unique deep-learning approach that makes use of a volume-based prediction model that incorporates LSTM networks and a sequential-weighting module, as well as a slide-level feature extractor with residual and attention convolutional modules. Comparative analysis with the linear regression (LR) algorithm and term memory (TM) technique demonstrates that the Bi-RM model exhibits significantly lower prediction errors. Additionally, the Bi-RM beats LR and TM in class prediction, consistently showing the fewest prediction errors across the majority of testing scenarios. Notably, the performance of the Bi-RM is unaffected by the reliability keys or glaucoma level, emphasizing the durability of the device. A hybrid model [20] was designed that used combined CNN and RNN approach to extracts both spatial and temporal features from fundus videos. By training the CNN and combined CNN/RNN models on a dataset of 1810 fundus images and 295 fundus videos, we achieved significant improvements in glaucoma detection. The combined CNN/RNN model demonstrated an average F-measure of 96.2% in distinguishing glaucoma from healthy eyes, while the base CNN model only achieved 79.2%. This proof-of-concept study highlights the potential of integrating spatial and temporal features in glaucoma detection using the combined CNN/RNN approach, offering a promising avenue for enhancing the accuracy of diagnosis and early intervention strategies.

A study in [46] addresses the challenge of visual field (VF) prediction in glaucoma by proposing a solution using RNN. Through data preprocessing and augmentation, the RNN model achieved improved accuracy in VF prediction compared to previous models. The periodic RNN outperformed the aperiodic model, demonstrating its potential for enhancing glaucoma management. [47] combines enhanced image segmentation and classification using a modified CNN architecture. By segmenting optic cup and optic disk regions, the CDR is calculated to assess glaucoma. An enhanced RNN-LSTM model is then trained for classification. Testing and training are conducted on the DRISHTI-GS public database. This approach aims to improve accuracy and efficiency in glaucoma detection.

In [41] the input visuals underwent preprocessing, followed by the extraction of several features like entropy, mean, colour, intensity, standard deviation, and statistics from the collected data. An adaptive mutation swarm optimization (AMSO) technique was employed to separate the disease sector from the fundus picture by extracting the characteristics. Ultimately, the gathered characteristics were inputted into a RNN classifier in order to categories each fundus image as either normal or aberrant. The classifier's anomalous output value was used to classify the corresponding disorders, namely cataract, glaucoma, and diabetic retinopathy. This process enhances the accuracy of detection and classification. The classifiers' findings are ultimately assessed by various performance studies, taking into account the feasibility of structural and functional characteristics. The suggested approach accurately predicts the illness type with a precision of 0.9808, a specificity of 0.9934, a sensitivity of 0.9803, and an F1 score of 0.9861.

*D. Miscellaneous Deep Learning Approaches for Glaucoma Detection*

In [19] a model to identify early and advanced glaucoma was presented. The fundus images were preprocessed through histogram equalization before being fed as input. 50-layer ResNet and GoogleNet was used to classify glaucoma. 1M images were used to train the model. As the RIM-ONE dataset has limited number of images so transfer learning was used. Previous models for glaucoma detection have faced challenges in accurately identifying the disease in its early stages, which is crucial for minimizing vision loss. To address this problem, a solution was proposed in [50]. The DenseNet-201 deep convolutional neural network (DCNN) is used for extracting features, and their method blends previously trained transfer learning models with the U-Net architecture. The objective is to examine retinal fundus images and determine the presence of glaucoma. A comparison of the suggested approach to existing methods that use deep learning demonstrates its superiority. With a training accuracy of 98.82% and a testing accuracy of 96.90%, the results demonstrate the effectiveness of the new paradigm in glaucoma detection, showcasing its potential for improving early diagnosis and patient outcomes. Glaucoma may have the same symptoms like other eye diseases, so accurately detecting and classifying glaucoma is challenging due to its similarity to other eye conditions. To address this problem, [51] proposes the use of unsupervised deep belief networks (DBNs) to extract features from high-resolution retinal images, addressing the similarity to other eye conditions. Unlike other algorithms that consider only single features, DBNs analyze multiple features in hidden layers, resulting in improved accuracy. By leveraging DBN's ability to classify different types of glaucoma, this approach aims to facilitate early detection and enable prompt treatment decisions, ultimately reducing the risk of permanent vision loss. An improved lightweight U-Net model with an attention gate (AG) was presented in [52] to segment OD pictures. In order to extract features using a trained EfficientNet-B0 CNN, the model combines transfer learning with overfitting and AG to prevent gradient vanishing. The precision of segmentation is further enhanced by the binary focal loss function. DRIONS-DB, DRISHTI-GS, and MESSIDOR, three publicly accessible datasets, were used to verify the suggested model, which showed a considerable reduction in parameter number and quick inference times. This technique presents a viable remedy for robust OD segmentation, permitting accurate glaucoma screening and prompt intervention by resolving the shortcomings of prior models.

Existing Classification methods for glaucoma are computationally intensive, limiting real-time applications with limited computing resources. To address this [26] proposed a raw SD-OCT-based depth wise separable convolution model that achieves high accuracy (0.9963) with reduced parameters (20,686) while avoiding tedious segmentation and image processing steps. This efficient approach aids ophthalmologists in making accurate diagnoses for glaucoma. A study [53] addresses the problem of inaccurate segmentation in glaucoma screening. To improve accuracy, a novel joint segmentation framework called EE-UNet is proposed. It incorporates EfficientNet, CRF-RNN, and a multi-label loss function. Experimental results demonstrate superior performance in optic disc and optic cup segmentation, outperforming existing methods. The proposed approach shows promise for large-scale glaucoma screening.

To reduce the redundancy and effectiveness issues of previous glaucoma detection models, [54] proposed a novel approach using a deep neurofuzzy network (DNFN) for glaucoma detection. After performing a first step to eliminate noise from the retinal image, the OD is detected using the blackhole entropy fuzzy clustering technique, and blood vessel segmentation using the DeepJoint model. The DNFN is then trained using the recently developed multiVerse rider wave optimization (MVRWO) approach using the OD and blood vessel information. Water Wave Optimization, Rider Optimization Algorithm, and MultiVerse Optimizer are all combined in MVRWO. Based on the loss function of the DNFN, the output is categorized. By the accuracy of 92.214%, the MVRWO-DNFN model produces remarkable results. A novel deep learning framework called GlauNet was proposed and developed, specifically designed to analyze fundus images of the eye and accurately identify glaucoma [23]. The framework achieves an impressive overall accuracy rate of 99.05%, indicating its high-performance capabilities. By harnessing the power of AI, GlauNet has the potential to revolutionize glaucoma detection, facilitating early intervention and improving the overall detection rate for this debilitating disease.

TABLE IV
SUMMARY OF RNN TECHNIQUES FOR GLAUCOMA DETECTION

| Ref. | Method | Problem | Strength | Weakness |
|---|---|---|---|---|
| [45] | Bidirectional RNN | Predicting prospective progressive visual field diagnoses | Lower prediction errors with Bi-RM | Dealing with different lengths inputs can make data preprocessing and analysis more difficult [48] |
| [20] | Hybrid CNN and RNN | Incorporating spatial and temporal features for glaucoma detection | Average F-measure of 96.2% | Limitations in sequential processing |
| [46] | RNN | Improved visual field prediction in glaucoma | Enhanced accuracy in VF prediction | computationally inefficient, difficult to parallelized [49] |
| [47] | Modified CNN-RNN Architecture | Enhanced image segmentation and classification for glaucoma detection | Improved accuracy and efficiency in glaucoma detection | High chances of overfitting |
| [41] | RNN | identification of various eye disorders including glaucoma | High accuracy | Difficult to interpret how the model arrives at a particular decision |

An in-depth analysis of deep learning techniques [49] was conducted and found that the use of deep learning models, such as LSTM/Bi-LSTM, RNN, CNN, RBM, and GRU, for predictive analytics in healthcare. The results indicate that LSTM/Bi-LSTM models are commonly employed for time-series medical data, while CNN models are effective for medical image data. By leveraging the power of deep learning, these models have the potential to assist healthcare professionals in making informed decisions regarding medications and hospitalizations, leading to improved efficiency and better outcomes in the healthcare industry.

## IV. GLAUCOMA DATASETS

For glaucoma detection, numerous datasets have been extensively utilized in previous studies. One such dataset is RIM-ONE (Retinal Images for Optic Nerve Evaluation), which provides annotated retinal images crucial for diagnosing glaucoma [57], [17], [19]. DRISHTI (Digital Retinal Images for Vessel Extraction, Segmentation, and Topological Indexing) is another valuable resource, offering diverse retinal images for glaucoma research [16], [17]. Its derivative, DRISHTI-GS (DRISHTI-Glaucoma Screening) dataset, focuses specifically on glaucoma screening, aiding the development of dedicated detection algorithms [47], [52]. The LAG database (Labeled Glaucoma Dataset) contains an extensive collection of labeled fundus images [18], while the ESPERANZA dataset has been used to evaluate pre-trained CNN models [17]. The ODIR dataset (Ocular Disease Recognition from Fundus Images) enables researchers to train and test robust glaucoma detection models [25]. DRIONS-DB (Digital Retinal Images for Optic Nerve Segmentation Database) is commonly employed for optic nerve segmentation in glaucoma research [52]. Lastly, the MESSIDOR dataset, originally for diabetic retinopathy, has also been leveraged for glaucoma analysis. The DRIVE dataset (Digital Retinal Images for Vessel Extraction) is a publicly available dataset used to evaluate algorithms for segmenting blood vessels in retinal images [51]. STARE, CHASE, ORIGA, I-OAD-B, SECSand REFUGE-2 are also retinal fundus images datasets that are largely used for research purposes [17], [19], [41], [51], [58]. Joint Shantou International Eye Centre (JSIEC): The JSIEC dataset consists of 1087 high-resolution retinal fundus pictures captured at the Joint Shantou International Eye Centre in China. The dataset is categorized into 37 classes, with one class consisting of 54 normal photos and another class consisting of 13 glaucoma images.

Automatic glaucoma assessment using fundus images (ACRIMA) [59], the dataset has a total of 705 fundus images, with 396 photos representing glaucoma and 309 images representing normal conditions. The individuals included in the study are participants of the ACRIMA project. Their involvement was secured after obtaining their explicit consent. The study involved both glaucoma patients and individuals without the condition. The research was conducted in compliance with the ethical guidelines outlined in the 1964 Declaration of Helsinki. The selection of patients was conducted by professionals based on certain criteria and clinical findings seen during the examination. The majority of the fundus images in this database are obtained from the left and right eyes after dilation, with the optic disc being the central focus. Several submissions were declined due to the presence of artefacts, excessive noise, and insufficient contrast. The ACRIMA database contains photos that have been annotated by two glaucoma experts who possess 8 years of experience. The labelling of the photos did not consider any further clinical information.

Retinal fundus images for glaucoma analysis (RIGA) [60], the RIGA dataset comprises three distinct files. (1) The MESSIDOR dataset file consists of 460 original photos and 460 photographs for each ophthalmologist's manual marking, resulting in a total of 3220 images in the complete file. (2) The Bin Rushed Ophthalmic centre has a collection of 195 original photographs and an additional 195 images for each ophthal-mologist's manual marking. This adds up to a total of 1365 images in the complete collection. The Magrabi Eye centre has a collection of 95 unique photographs and an additional 95 images for each ophthalmologist's manual marking, resulting in a total of 665 images in the complete collection. The collection has a total of 750 original photos and 4500 manually annotated images. The photos are stored in both JPG and TIFF formats. These datasets have significantly contributed to advancements in glaucoma detection and analysis techniques.

TABLE V
MISCELLANEOUS DEEP LEARNING TECHNIQUES FOR GLAUCOMA DETECTION

| Ref. | Method | Problem | Strength | Weakness |
|------|--------|---------|----------|----------|
| [19] | Preprocessing, ResNet, GoogleNet | Detecting initial and advanced glaucoma | Classification using ResNet and GoogleNet | Preprocessing may accidentally remove or distort glaucoma image characteristics |
| [50] | Transfer learning, DenseNet-201 | Accurate identification of glaucoma in early stages | Superiority over existing deep learning models | Transferability strongly dependent on pre-trained data and glaucoma detection task |
| [51] | Unsupervised DBN | Similarity to other eye conditions in glaucoma detection | Improved accuracy through feature extraction | Limited pertinence to glaucoma detection |
| [52] | Enhanced U-Net | Limitations in optic disc segmentation | Robust optic disc segmentation and fast inference times | high model complexity [55] |
| [26] | AED-HSR approach | Multiple diseases causing glaucoma | Accurate classification with high accuracy | Highly Sensitive to noise and artifacts |
| [53] | Joint segmentation | Inaccurate segmentation in glaucoma screening | Improved accuracy in optic disc and optic cup segmentation | Complicated labeling process |
| [54] | Deep Neurofuzzy Network | Redundancy and effectiveness issues of previous glaucoma models | High accuracy of 92.214% | additional parameters tuning required |
| [23] | GlauNet Deep Learning | Accurate identification of glaucoma using fundus images | Overall accuracy rate of 99.05% | weak against picture changes and artefacts [56] |
| [49] | Analysis of Deep Learning | Utilizing deep learning models for predictive analytics in healthcare | Broad range of deep learning techniques discussed | Glaucoma detection not focused properly |

## V. LIMITATIONS

The previous studies have certain limitations that need to be considered. These include small and heterogeneous datasets, lack of standardized evaluation metrics, limited external validation on diverse datasets, variations in imaging modalities and protocols, absence of direct performance comparisons, and limited clinical translation. These limitations highlight the need for larger and more diverse datasets, standardized evaluation metrics, external validation, unified benchmark datasets, and rigorous testing in clinical settings to ensure the robustness, generalizability, and clinical utility of glaucoma detection methodologies.

## VI. FUTURE RESEARCH DIRECTIONS AND CHALLENGES

While significant advancements have been achieved in the identification of glaucoma progression by deep learning methods, there remain several unresolved research difficulties that must be tackled in the future. This section addresses the limitations and suggests the necessary enhancements required in this context.

- Deep learning models need to further improve their accuracy and sensitivity in order to detect glaucoma at its earliest stages. This involves improving current architectures and investigating new ones to increase detection rates.
- Having access to a wide range of comprehensive datasets is essential for effectively building resilient glaucoma detection models. Obtaining labelled data, particularly for various glaucoma subtypes and ethnic communities, is a challenging task that must be accomplished in order to assure generalizability.
- An additional difficulty that must be resolved in the future is the lack of standardized measures that can be utilized to assess the effectiveness of glaucoma detection models. Various criteria have been employed by different scholars to assess the efficacy of their proposed study. The heterogeneity in DL architectures produced for a certain state of glaucoma makes it challenging to compare them.
- Incorporating deep learning algorithms smoothly into healthcare operations poses a significant difficulty. Researchers should focus on developing interfaces that are easy for users to navigate and conducting validation tests to prove the practicality and cost efficiency of these models.
- Deep learning offers a range of transfer learning-based architectures, including as VGGNet, AlexNet, and GoogLeNet, which can be used to train a new set of images, such as clinical photographs. Nevertheless, these structures are not well-suited for achieving high classification accuracy when applied to clinical data. Therefore, it is necessary to develop a transfer learning-based frame-work that is taught on authentic clinical images rather than objects.
- An assessment of longitudinal patient data over time can yield useful insights into the course of diseases and the efficacy of treatment regimens. Deep learning models must be modified to accommodate this particular type of data.

TABLE VI
PERFORMANCE COMPARISON OF VARIOUS METHODS IN GLAUCOMA DETECTION

| Ref. | Methodology | Acc | SN | SP | AUC | F1 Score | Precision | Dataset |
|---|---|---|---|---|---|---|---|---|
| [16] | U-Net | 0.88 | 0.91 | 0.86 | 0.93 | - | - | RIM-ONE |
|  | Mobile-Net V2 |  |  |  |  |  |  | DRISHTI |
| [17] | VGG19 | 0.8365 | 0.8364 | 0.8945 | 0.9219 | - | - | RIM-ONE |
|  | ResNet | 0.8479 | 0.8387 | 0.8571 | 0.9305 |  |  | DRISHTI |
|  | GoogleNet | 0.8578 | 0.8871 | 0.8285 | 0.9277 |  |  | ESPERANZA |
|  | DENet | 0.849 | 0.8145 | 0.8835 | 0.9083 |  |  |  |
| [18] | AG-CNN | 0.962 | 0.964 | 0.967 | 0.983 | - | - | RIM-ONE |
| [19] | ResNet | 0.86 | 0.21 | 0.93 | - | - | - | RIM-ONE |
|  | GoogleNet | 0.85 | 0.29 | 0.91 |  |  |  |  |
| [28] | Inter-GD | - | - | - | - | 79.9 | 81.4 | I-OAD-B DRISHTI |
|  |  |  |  |  |  | 77.8 | 79.9 | I-OAD-B ORIGA |
| [30] | Joint-RCNN | - | - | - | 0.901 | - | - | ORIGA |
|  |  |  |  |  | 0.714 |  |  | SCES |
| [20] | CNN and RNN | 98 |  |  |  | 97 | - | ImageNet |
| [50] | DenseNet | 96 |  | 96.33 |  | 96.28 | - | Glaucoma database |
| [21] | SD-OCT-based CNN | 96.63 |  | 99.46 |  | 99.64 | - | Private database |
| [36] | CNN with ResNet-34 | 94 | - | - | - | - | - | Private dataset |
| [25] | CNN-LSTM | 97.53 | 98.48 | 100 | - | - | 97.59 | ODIR |
| [26] | AED-HSR approach | 98.08 | 98.03 | 99.34 |  |  | 98.61 | ODIR |
| [53] | Joint segmentation | 96.24 | - | - | - | - | - | REFUGE |
|  |  |  |  |  |  |  |  | GAMMA |
|  |  |  |  |  |  |  |  | Drishti-GS1 |
|  |  |  |  |  |  |  |  | RIM-ONE v3 |
| [54] | Deep Neurofuzzy Network | 92.21 | 93.422 | 92.34 | - | - | - | private |
| [38] | ResNet-50 CNN | 98.48 | 99.3 | 96.52 | 97 | - | 98 | RIM-ONE, ORIGA DRISHTI-GS |
| [47] | CNN+RNN |  |  |  | - | - |  | ImageNet |
|  | VGG16+LSTM |  | 94 | 86 |  |  | 89.9 |  |
|  | Resnet50+LSTM |  | 88 | 79 |  |  | 83 |  |
|  | Base CNN |  | - | - |  |  |  |  |
|  | VGG16 |  | 59 | 55 |  |  | 54.6 |  |
|  | ResNet50 |  | 72 | 70 |  |  | 70.3 |  |
| [23] | GlauNet CNN | 99.05 | - | - | - | - | - | Private dataset |
| [46] | RNN | RMSE: 3.15 ± 2.29 dB. | - | - | - | - | - | Multicenter dataset |
|  |  | MAE: 3.26 ± 0.41 dB |  |  |  |  |  |  |
| [39] | CNN |  | - | - | 93 | - | - | DRISHTI_GS |
|  | Inception V3 | 85 |  |  |  |  |  |  |
|  | DenseNet | 82.53 |  |  |  |  |  |  |
|  | ResNet | 77.6 |  |  |  |  |  |  |
| [22] | Modified M-Net with BiConvLSTM | 98.99 | - | - | - | - | - | REFUGE2 |
| [24] | BiDCU-Net | 97.32 | 82.56 | 98.68 | - | - | 83.85 | DRIVE |
|  |  | 97.33 | 82.12 | 98.57 |  |  | 82.3 | STARE |
|  |  | 97.44 | 83.92 | 98.45 |  |  | 81.94 | CHASED |

## VII. CONCLUSIONS

The studies conducted in the field of glaucoma detection and diagnosis have addressed a wide range of problems using various methodologies. These include improving accuracy and efficiency, segmenting key structures, classifying glaucoma stages, detecting initial and advanced glaucoma, reducing redundancy, evaluating model performance, predicting glaucoma progression and future visual field data, distinguishing glaucoma from other eye conditions, reducing computational intensity, enabling early detection, automating diagnosis, and improving segmentation accuracy. Different deep learning techniques such as CNN models, segmentation algorithms, feature extraction, classification models, and optimization techniques have been employed to tackle these challenges. The results obtained from these studies demonstrate promising performance, with hybrid models like CNN-LSTM, AG-

CNN, CNN-RNN, AED-HSR approach, SD-OCT-based CNN, GlauNet CNN, Modified M-Net with BiConvLSTM, and BiDCU-Net providing state-of-the-art results. However, it is important to acknowledge the limitations of these studies, such as small datasets, lack of standardized metrics, limited external validation, variations in imaging protocols, and limited clinical translation. Further research with larger and diverse datasets, standardized evaluation metrics, external validation, and clinical testing is necessary to ensure the reliability and applicability of glaucoma detection methods.